\documentclass[10pt,letterpaper]{article}

\usepackage{cogsci}

\cogscifinalcopy 

\usepackage{microtype}
\usepackage{amsmath}
\usepackage{soul}
\usepackage{pslatex}
\usepackage{apacite}
\usepackage{float} 

\usepackage{graphicx}
\usepackage{subcaption}
\graphicspath{{./figs/}}
\usepackage[table]{xcolor}
\usepackage{dirtytalk}
\usepackage{natbib}
\usepackage{hyperref}
\usepackage{booktabs}
\usepackage{caption}
\usepackage{units}
\usepackage{threeparttable}
\usepackage{tabularx}

\newcolumntype{Y}{>{\centering\arraybackslash}X}
\title{Selective imitation on the basis of reward function similarity}
 
\author{{\large \bf Max Taylor-Davies (s2227283@ed.ac.uk)} \\
        Edinburgh Centre for Robotics \\
        University of Edinburgh
    \AND {\large \bf Stephanie Droop (stephanie.droop@ed.ac.uk)} \\
        Institute for Language, Cognition and Computation \\
        University of Edinburgh
    \AND {\large \bf Christopher G. Lucas (c.lucas@ed.ac.uk)} \\
        Institute for Language, Cognition and Computation \\
        University of Edinburgh
}

\begin{document}

\maketitle

\begin{abstract}
Imitation is a key component of human social behavior, and is widely used by both children and adults as a way to navigate uncertain or unfamiliar situations. But in an environment populated by multiple heterogeneous agents pursuing different goals or objectives, indiscriminate imitation is unlikely to be an effective strategy---the imitator must instead determine who is most useful to copy. There are likely many factors that play into these judgements, depending on context and availability of information. Here we investigate the hypothesis that these decisions involve inferences about other agents' reward functions. We suggest that people preferentially imitate the behavior of others they deem to have similar reward functions to their own. We further argue that these inferences can be made on the basis of very sparse or indirect data, by leveraging an inductive bias toward positing the existence of different \textit{groups} or \textit{types} of people with similar reward functions, allowing learners to select imitation targets without direct evidence of alignment. 

\end{abstract}

\section{Introduction}
\label{intro}
The complexity and variety of the real world is such that we often find ourselves required to act in new environments or unfamiliar scenarios. When determining what actions to take in such situations, we could follow an exploratory approach, trying lots of different behaviors until we achieve our desired result. But this will likely prove inefficient, and in some circumstances may even be dangerous or intractable. Imitation provides an attractive alternative---if we can observe other agents in our environment, perhaps we can learn to follow their example. Children are quick to learn by imitating others, sometimes faithfully and sometimes selectively \citep{howard2015infants}, although their strategies are still not completely understood \citep{over2013social}. Selectivity makes sense in the context of bounded resources: in an environment populated by diverse agents, not all will be equally useful to imitate. Indeed, copying an unknown agent may actually be counterproductive: the stranger may be eating something intolerably spicy, or be able to navigate terrain, like deep water, that we cannot. Therefore, any would-be imitator must engage in some form of selection or filtering, identifying which agent(s) to imitate and which to ignore (or in some cases electing to avoid imitation altogether). This process likely relies on many different factors, and is heavily context-dependent. We will first give a brief overview of existing research that attempts to map out these factors, before considering one in particular that we believe to be underexplored. 

\subsection{Social learning strategies}
\label{sls}
A substantial body of empirical and theoretical research into the field of \textit{social learning strategies}, primarily from an evolutionary perspective, has identified a range of rules or heuristics used by both humans and other animals to guide their selection of imitation targets \citep{Rendell2011}. One strategy, observed in stickleback fish \citep{Kendal2009} and humans alike \citep{Zmyj2010}, is to copy the agents that are observed receiving the highest payoff within the domain of interest (such as locating sources of food). Other strategies rely on less task-specific characteristics of potential models. For instance, stickleback fish also prefer to copy the actions of larger demonstrators \citep{Duffy2009}, and chimpanzees prefer to copy older individuals and those occupying higher social rank \citep{Horner2010}. Human children also imitate based on age and social status, preferring to copy adults rather than their same-age peers, even when those peers possess better domain knowledge \citep{Wood2012}. Children additionally take into account the familiarity of potential model agents, placing higher trust in information provided by a more familiar teacher \citep{Corriveau2009}.  

\subsection{Inferring others' reward functions}
\label{tom}
Heuristics such as ``pick the older/more successful/more familiar person'' likely do not capture the full picture of how and when people choose who to imitate. For instance, they do not entail any direct reference to the cognitive or mental states of either imitator or model. It is well established that people develop, from a young age, the ability to reason about the hidden mental states of others from observations of their behavior---typically referred to as Theory of Mind (ToM) \citep{Wellman1990TheCT, Fodor1992ATO, Repacholi1997EarlyRA, Onishi2005Do1I}. It is plausible that reasoning about the cognition of other agents could be useful in determining their suitability as potential imitation targets, and \citet{Heyes2016} has argued for the existence of a subset of social learning strategies (SLSs) that are explicitly metacognitive. An example of how an SLS may rely on ToM is given by \citet{Diaconescu2014}. When faced with a repeated binary-choice task, and provided with advice from an advisor motivated either to help or mislead them, participants followed the advice to the extent that they believed their advisor wanted to be helpful. 

One key component of ToM concerns the determination of other agents' goals or reward functions. This inference process is usually invoked in the context of predicting the behavior of another agent in order to manage some sort of interaction with them, but it may also play a role in selecting which agent(s) to imitate. In recent years, efforts have been made to capture this process using computational models. \citep{Lucas2014, JaraEttinger2015ChildrensUO}. One promising approach has been to develop models based on the idea of \textit{inverse planning} \citep{baker2009action, Pantelis2014, baker2017rational, zhi2020online} or \textit{inverse reinforcement learning} \citep{Ramachandran2007BayesianIR, Ziebart2008MaximumEI}, where generative models of (approximately) rational behavior are inverted to produce inferences of mental states or reward functions from observed actions. These models are often evaluated using simple 2D environments, referred to as ``gridworlds". A gridworld usually contains a small fixed set of possible goal items or states, and agents within the gridworld are limited to an action space that consists only of movement along the four cardinal directions. While representing a significant simplification relative to naturalistic settings, their bare-bones structure allows for clear and unambiguous presentation of goal-directed agent behavior, as well as enabling tractable modelling. Although our aim in this paper is not in explicitly evaluating or modelling people's ability to infer reward functions \emph{per se}, this line of research inspires both our choice of experimental paradigm and the predictions we make in the following section. We believe that simple gridworld environments are well-suited as a setting for exploring how people \textit{use} the inferences they make about other agents' reward functions to guide their own behavior---something which, to our knowledge, they have not previously been used to investigate. 

\subsection{Imitation on the basis of reward function inference}
\label{questions}
Knowing that people engage both in selective imitation and in inferring the reward functions of others in their environment, we can pose a simple question: do these inferences inform the selection of imitation targets? More specifically, do people selectively imitate those in their environment they believe have reward functions that are similar in some sense to their own (\textbf{Question 1})? In the following sections of this paper, we attempt to test this proposal. Extending this simple question, we also suggest and investigate two possible ways in which people might \textit{generalize} their decisions about who to imitate: \textit{goal generalization} and \textit{agent generalization}. First, when placed in a new context where they have limited or no information about \textit{their own reward function}, do people continue to imitate those who in a previous context were judged as having a reward function similar to their own (\textbf{Question 2})? Second, can people take advantage of correlations between the reward functions of different agents to select imitation targets without comparing their own reward function directly (\textbf{Question 3})? To investigate these questions, we conduct an experiment using navigation-like tasks in an online gridworld environment. In the following section we describe in more detail the structure of the experiment, and outline the predictions made. 

\section{Experiment}
\label{experiments}

\begin{figure*}
    \centering
    \includegraphics[width=14cm]{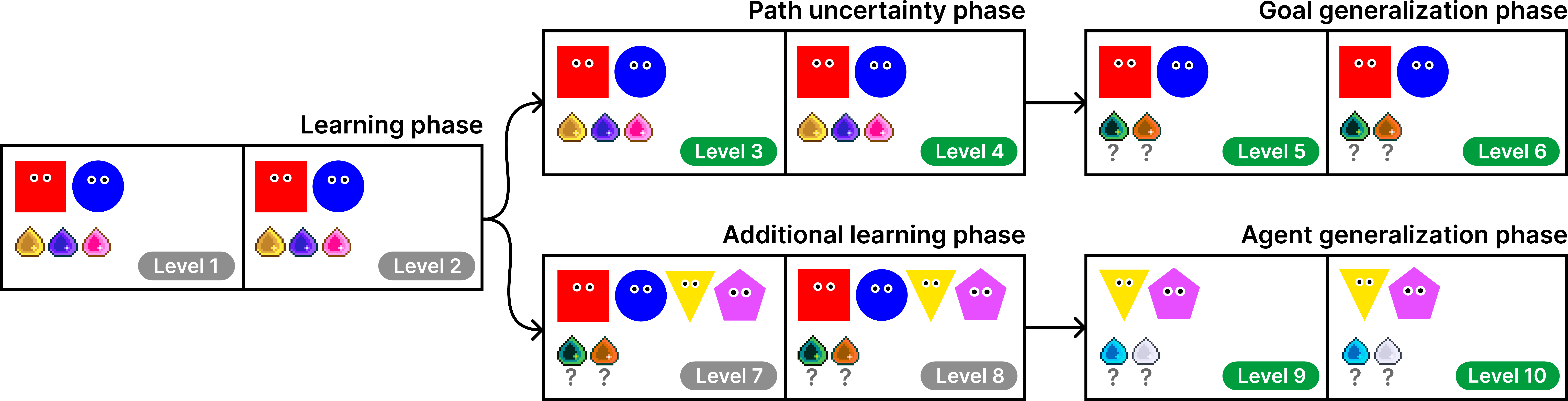}
    \caption{A diagram of the experiment structure, showing the progression of levels and indicating which agents and gems were present in each. Green labels indicate levels included in the analysis; grey labels indicate learning/training levels. Gems whose value were unknown to the participant are highlighted with a question mark.}
    \label{fig:experiment}
\end{figure*}

To test the three questions posed in the previous section, we conduct an online experiment, in which participants navigate a series of 2D gridworld environments (``levels") and score points by collecting colored ``gems". Each level is populated by a set of simulated agents, visible to participants, that collect gems to maximize their own fixed reward functions. By applying restrictions to the availability of information in certain levels, we incentivize participants to imitate agents' behavior, with their choice of imitation target reflected in the path they choose to take through the level. We manipulate two factors between participants. Participants are assigned to either the \textit{path+goal} condition (addressing \textbf{Questions 1-2}) or the \textit{agent} condition (addressing \textbf{Question 3}); in addition, they are randomly assigned one of two fixed reward functions ($\mathbf{r}_1$ or $\mathbf{r}_2$), which determine the mapping from gem colors to points. 

Following an unscored practice level, the experiment is divided into distinct phases that each consist of two levels (see Fig. \ref{fig:experiment} for a visual schematic). The phases completed by each participant depend on which condition they are assigned. To begin with, all participants complete the \textbf{learning phase}, which provides evidence of the reward functions of Agents 1 and 2 with respect to gems A,B,C. In the \textit{path+goal} condition, the \textbf{learning phase} is followed by the \textbf{path uncertainty phase}. In this phase, the environment becomes partially observable (see Fig. \ref{fig:levels}), such that the participant knows the value of all gems available but not where each is located, making imitation the optimal strategy. This phase is followed finally by the \textbf{goal generalization phase} - while the environment in this phase is fully observable, participants still face uncertainty since all the gems present are new and have unknown value.

In the \textit{agent} condition, participants proceed from the first \textbf{learning phase} to the \textbf{additional learning phase}. In this phase, two new agents (Agents 3 and 4) are introduced, and the participant receives evidence of how the new agents relate to the original agents (Agents 1 and 2) in terms of reward function. The gems in this phase are also new; furthermore, the participant does not collect any gems themselves (remaining a passive observer). This means that they receive no \textit{direct} evidence about how the new agents' reward functions relate to their own. Finally, this phase is followed by the \textbf{agent generalization phase}, in which the original agents are removed, and the participant imitates one of the new agents in a choice between gems of unknown value. 

\subsection{Predictions}
\label{predictions}
The predictions made for the different phases were as follows:

\begin{enumerate}
    \item (\textit{Path uncertainty phase}) when people cannot see the location of the gems, nor which gems the agents collected, but can only see which direction each agent travelled in, they should choose to follow the direction of the agent which previously collected gems that maximizes their reward function. 
    \item (\textit{Goal generalization phase}) when people do not know the values of the gems available, they should generalize their previous choice of imitation target to the new set of options (imitating the same agent as before). 
    \item (\textit{Agent generalization phase}) when faced with a choice of unfamiliar agents to imitate, people should use evidence of correlations between agents' reward functions to identify and imitate the agent most likely to be aligned with their preferences. More specifically, suppose that they judge that Agent 1 has a similar reward function to them over one set of gems, and then see that Agent 3 makes the same choices as Agent 1 over a second set of gems. If a participant believes other agents' preferences to be correlated (e.g. because agents belong to some set of latent groups or types) then when faced with a third set of gems they should imitate Agent 3.
\end{enumerate}

\section{Methods}
\label{methods}

\subsection{Participants}
\label{ppts}
We recruited 150 UK-based adults through the online platform \href{https://www.prolific.co/}{Prolific}. Participants were paid £1.05 for taking part, plus a bonus of £0.01 for every 5 points they scored (mean £1.34, min £1.15, max £1.43). The experiment lasted 7m 55s $\pm$ 4m 27s. Each participant was randomly assigned with uniform probability to either the \textit{path+goal} ($N = 72$) condition or the \textit{agent} condition ($N = 78$).

\subsection{Stimuli}
\label{stim}

\begin{figure}
    \centering
    \includegraphics[width=8.5cm]{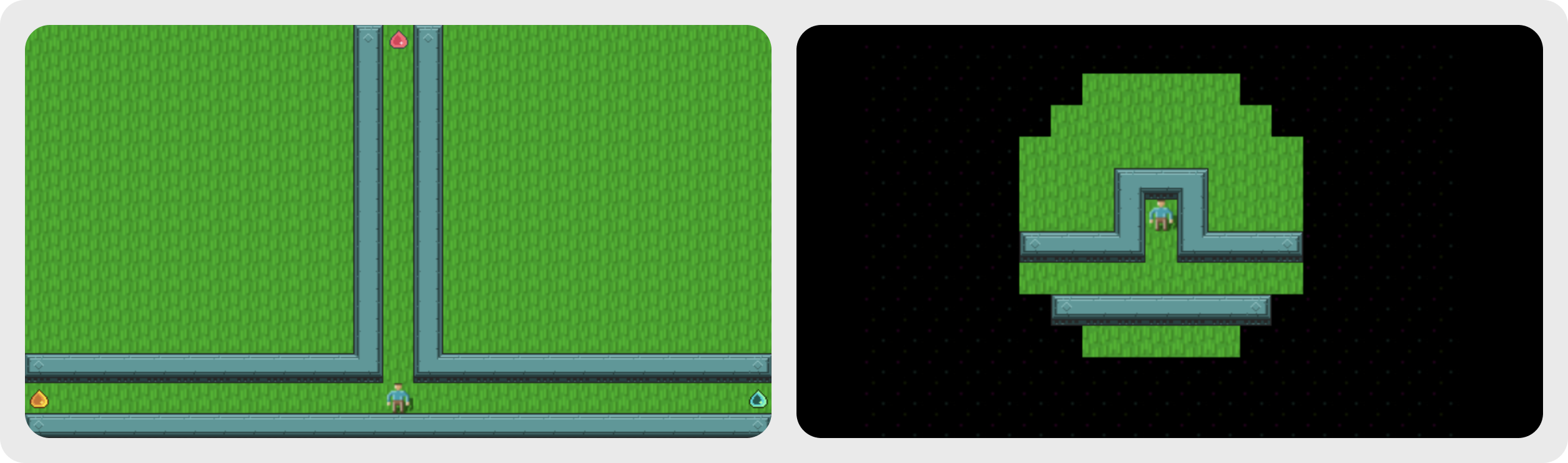}
    \caption{Examples of fully observable (left) and partially observable (right) gridworld levels.}
    \label{fig:levels}
\end{figure}

The experiment took place within an online 2D gridworld environment created using the GriddlyJS framework \citep{Bamford2022} and hosted on a custom web platform. Participants completed a number of levels within this environment, each containing a number of colored gems out of a total set of seven gems $\{A, B, C, D, E, F, G\}$. Participants complete a level by collecting any gem, but the number of points obtained varies for each gem depending on their assigned reward function. Participants were informed at the beginning of the experiment of the values of only the first three gems. Each level was either fully observable or partially observable; within a partially observable level, only gridworld tiles within a certain radius of the player avatar's current position are visible (Figure \ref{fig:levels} shows examples of both level type).

\subsection{Reward functions}
\label{reward-functions}
Across both the \textit{path+goal} condition and \textit{agent} condition, each participant was assigned one of two different reward functions ($\mathbf{r}_1$ or $\mathbf{r}_2$), which remained fixed throughout the experiment and controlled how many points were obtained for each color of gem. To encourage efficient trajectories, both reward functions also imposed a fixed cost of 1 point for every step taken in the environment. A participant's reward function thus completely determined the optimal trajectory for any given level. Each agent observed within the environment also had one of these two reward functions ($\mathbf{r}_1$ for Agents 1 and 3, $\mathbf{r}_2$ for Agents 2 and 4), and always executed the corresponding optimal trajectory. For the sake of brevity, we will use the terms \textit{aligned} and \textit{misaligned} to refer to agents with the same or different reward functions, respectively. For example, for a participant with reward function $\mathbf{r}_1$, Agents 1 and 3 were aligned, while Agents 2 and 4 were misaligned.

\subsection{Procedure}
\label{procedure}
Following an unscored practice level, participants completed a sequence of six levels in the gridworld environment, as described in \nameref{experiments}. The levels presented to each participant depended on whether they were assigned to the \textit{path+goal} condition or the \textit{agent} condition (see Fig. \ref{fig:experiment}). At the beginning of a level, the participant watched a sequence of prerecorded trajectories, each showing a different agent completing the level by navigating to one of the available gems (determined by their reward function). Agents were represented using stylised geometric avatars (see Fig. \ref{fig:experiment}). After watching these demonstrations, the participant completed the level themselves by using the arrow keys on their keyboard to move their own avatar around the gridworld to a gem. Their trajectory for each level was recorded as a sequence of 2D coordinates. At the end of the experiment, after completing all assigned levels, participants were asked to supply a short explanation (minimum 100 characters) for the choices they made.

\subsection{Analysis}
\label{analysis}
The trajectory of each participant was recorded for each level as a sequence of coordinates. For each of the levels included in the analysis there were two demonstrator agents shown; the trajectories of these agents were compared to the trajectory recorded for the participant. Letting $\tau_p$ and $\tau_j$ represent the trajectories of the participant and demonstrator agent $j$, respectively, we compute the \textbf{similarity} as  

\begin{equation}
    s(\tau_p, \tau_j) = e^{-\frac{1}{T}\sum_{t=1}^Td(\tau_p^{(t)}, \tau_j^{(t)})}
\label{eq:similarity}
\end{equation}

where $d(\tau_p^{(t)}, \tau_j^{(t)})$ gives the Euclidean distance between the two trajectories at timestep $t$. The function $s$ produces values that lie in the range $(0, 1]$, with a value of 1 indicating that the two trajectories are exactly identical. 

For each participant and level, we used the two similarity values (corresponding to the two agents) to compute a binary variable

\begin{equation}
\alpha_p = \arg\max_{j \in \{1,2\}}\{s(\tau_p, \tau_j)\}
\label{eq:argmax}
\end{equation}

indicating which agent participant $p$'s behavior was most similar to. This was used to perform (for each level) a single-tailed binomial test, with $n$ as the number of participants who completed the level, $k$ as the number who were more similar to the aligned agent, and a null hypothesis of $k/n = 0.5$. In addition, we performed a logistic regression to predict $\alpha_p$ from the participants' assigned reward function ($\mathbf{r}_1$ or $\mathbf{r}_2$).

\section{Results}
\label{results}

\begin{table*}
    \centering
    \begin{tabular}{ c c c c c c c }
        \toprule
        level & 3 & 4 & 5 & 6 & 9 & 10 \\
        $k / n$ & 63 / 72 & 63 / 72 & 50 / 72 & 55 / 72 & 52 / 78 & 41 / 78\\ 
        $p$ & $2.09 \times 10^{-11}$ & $2.09 \times 10^{-11}$ & $6.47 \times 10^{-4}$ & $4.07 \times 10^{-6}$ & .00328 & .411\\
        accuracy (\%) & 87.5 & 87.5 & 69.4 & 76.4 & 65.8 & 53.2 \\
        \bottomrule
    \end{tabular}
    \caption{The results of a single-tailed binomial test and a logistic regression. At each level, $n$ gives the number of participants that completed the level, and $k$ gives the number of participants that followed the aligned agent. $p$-values are computed based on the null hypothesis that people imitate the aligned and misaligned agents with equal probability (i.e. $k/n = 0.5$). The bottom row reports the per-level accuracy of a logistic regression model predicting imitation decisions (which agent a participant's trajectory was most similar to) from assigned reward functions ($\mathbf{r}_1$ or $\mathbf{r}_2$).} 
    \label{tab:results}
\end{table*}

\begin{figure*}[t!]
	\centering
    \includegraphics[width=17cm]{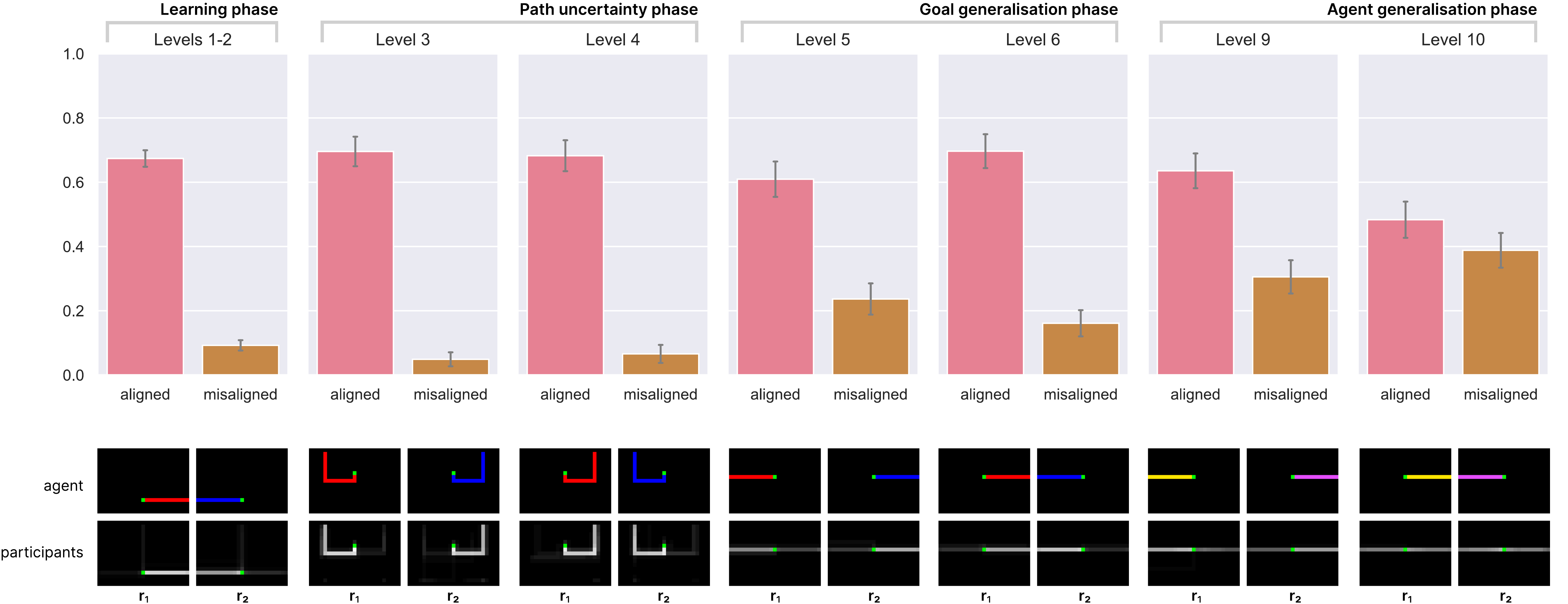}
	\caption{\textbf{Top}: mean similarity of participant trajectories to those of the aligned and misaligned agents for each level, computed following Equation \ref{eq:similarity}. Error bars represent standard error. \textbf{Bottom}: visualisation of recorded trajectories from participants and agents with  each reward function. Tiles with a lighter color were visited by a greater proportion of participants, and the starting location for each level is highlighted in green.}
	\label{fig:results}
\end{figure*}

\subsection{Exploratory analysis}
We performed exploratory analysis on participants' free text responses explaining their answers. For this, text responses were stripped of respondent or condition tags and manually coded for mention of various strategies. We observed that participants often mentioned following the agent with the preference for high scores, and only rarely explicitly mentioned transferring their allegiance to the agent who was like the first agent, which was the key experimental condition for Experiment 2, mentioned by 16 respondents out of 78. Those participants who did mention this strategy performed better at the task (average score 155) compared to those who did not pick up on this strategy (average score 135).

Unfortunately, despite the instructions and training phase, many participants did not understand the task or were not motivated to complete it successfully, as evidenced by mentioning they followed a random strategy (53 mentioned random). The participants who mentioned this performed worse at the task (average score 131) compared to those who did not ($N=98$, average score 156).

\subsection{Trajectory similarity}
For each level included in the analysis, we used Equation \ref{eq:argmax} as the basis of both a single-tailed binomial test and a logistic regression. Table \ref{tab:results} gives the results from both. During the \textit{path uncertainty phase}, participants were able to infer that when gems' locations were hidden, they could achieve the best outcome by following the agent which had previously been seen to favour their reward-maximising gem ($p < .001$, accuracy $= 87.5\%$). Furthermore, in the \textit{goal generalization phase}, participants were able to generalize this inference to novel environments containing only gems of unknown value ($p < .001$, accuracy $= 69.4\%$ in level 5; $p < .001$, accuracy $= 76.4\%$ in level 6). Finally, when presented with unfamiliar agents, and given evidence \textit{only} of the relation between new and old agents (and not directly between the new agents and themselves), participants in the \textit{agent generalization phase} were able to identify which of the new agents was more likely to be aligned with their own preferences and imitate them. This was seen in level 9 ($p = .00328$, accuracy $= 65.8\%$)---but in level 10 (the final level) participants were much more prone to explore, with imitation choices not significantly different from chance ($p = .411$, accuracy $= 53.2\%$). 

Figure \ref{fig:results} (top) shows, for each level across the three test phases, the mean trajectory similarity of participants to both aligned and misaligned agents. In the \textit{path uncertainty phase} (levels 3-4), the participants' behavior showed overwhelming similarity to that of the aligned agent. In the \textit{goal generalization phase} (levels 5-6), the trend is slightly weaker, but still we see significantly more similarity to the aligned agent. Finally, considering the \textit{agent generalization phase}, while for level 9 we see  significant favouring of the aligned agent, for level 10 the difference is within standard error, and so is not significant. This can likely be explained by participants having additional motivations not captured by their assigned reward functions, such as curiosity. Since the values of the gems in this phase were unknown, some participants in level 10 (the final level) will have explored in order to learn the value of whichever gem they didn't select in level 9. The visualisations of actual participant and agent trajectories in Figure \ref{fig:results} (bottom) provides an additional view into the same results.

\section{Discussion}
In this paper, we have explored the question of whether people's selection of imitation targets depends on inferences about their reward functions, and to what extent these selections are generalized. By conducting a behavioral experiment within a virtual gridworld environment, we found that when faced with a choice of imitation targets, people preferentially copy the behavior of an agent they judge to have a similar reward function to their own. Furthermore, we demonstrated evidence that people generalize these selections beyond the original context in which the inference was made, continuing to imitate the same agent's choices over a new set of options with unknown value. More interestingly, our results support the idea that people can extend their inferences not only to unknown items or goals, but also to \textit{unfamiliar demonstrators}. By using observed correlations between the reward functions of different agents, people are able to select appropriate imitation targets even without direct evidence of similarity to themselves.

\subsection{Imitating agents who share your reward function}
In any environment populated by heterogeneous agents pursuing different tasks or goals, imitating at random is unlikely to produce favourable results. In some cases, simple SLSs based on superficial factors like age or appearance will be sufficient to discriminate `good' targets from bad. But in other cases, these approaches will fall short. For example, we might have a setting where agents appear similar or identical in terms of explicit observable characteristics, but still vary significantly along dimensions that are important to determining their behavior. Alternatively, a highly dynamic and fast-changing environment could render more stable agent characteristics increasingly less informative. We argue that under conditions such as these, a metacognitive SLS based on reward function inference can provide a valuable alternative. By identifying and imitating agents whose behavior is directed towards the same task or goal that they are trying to accomplish, a learner can acquire behavior that is more likely to lead to outcomes satisfying their own reward function. 
Indeed, our results offer evidence for the existence of such a strategy: in agreement with Prediction 1 (see \nameref{experiments}), participants in the \textit{path uncertainty phase} showed an overwhelming preference for imitating the agent that they had previously been able to infer (during the \textit{learning phase}) shared their reward function over the available gems. However, it is also important to highlight certain limitations of the current experiment. For instance, as outlined in the \textit{Exploratory analysis} section, the free text responses provided by a number of participants indicated that they failed to understand the task. This suggests that the way the task is presented to participants should be improved in any future versions. Furthermore, 
a possible alternative explanation for participants' choices is that they judged the agents as having the same reward function but varying in how competent they were at satisfying it. In an attempt to preclude this, participants were instructed directly that agents were equally capable but could vary in their preferences for gems. However, future versions of this experiment should be designed in such a way as to explicitly distinguish between these two explanations.  

\subsection{generalization to unfamiliar domains}
In Prediction 2, we suggested that people can generalize these inferences to new contexts, where the relative value of all options is unknown. Our results from the \textit{goal generalization phase} support this prediction. Participants facing a novel choice between gems of unknown value were able to generalize their previous selection of imitation target. This has implications for how people handle situations involving uncertainty around \textit{their own reward function}. As an example, imagine ordering a meal in a foreign country whose food is completely unfamiliar to you. You may have no idea which of the many dishes you would like best, but if your friend has some experience with the cuisine, and you know from prior experience that you and your friend often have similar tastes, then you might assume that this similarity transfers to the new context and copy your friend's order. By doing this, you can reduce the uncertainty in your own reward function through selective imitation, effectively learning something about yourself by copying the behavior of someone else. Of course, even people with highly similar preferences don't agree on \textit{every} conceivable choice, and so this type of generalization will not always make sense. An interesting direction for future work would be to investigate the factors that determine people's cross-domain generalization of reward-function-based imitation. For instance, we might expect that people are more willing to generalize when the two domains have more in common, or when they are more familiar with the agent they're imitating. 

\subsection{generalization to unfamiliar agents}
While we have argued that inferring the reward functions of other agents in a heterogeneous environment can provide a useful basis for selecting imitation targets, there are reasons to believe that this picture is incomplete. Firstly, recovering a complete representation of an agent's reward function just from observations of their behavior is likely a sample-intensive process. In an environment populated by a large number of agents, it could therefore prove prohibitive to carry out this inference for every agent individually; especially when reward functions are high-dimensional. Furthermore, even in a complex environment that supports a large space of possible reward functions, the probability distribution over that space will likely be strongly peaked around a small number of points corresponding to common reward functions, with some variation. Encountering a new and unfamiliar agent, it is therefore likely that we can capture a reasonable approximation of their reward function just by assigning them to one of these points; in most cases this should be substantially more sample-efficient than trying to recover their reward function `from scratch'. 
As a first step towards investigating this idea, our results from the \textit{agent generalization phase} show that people can in fact judge the suitability of an unknown demonstrator by using evidence of correlations between different agents' reward functions. We suggest that this is enabled by an inductive bias that pushes people towards modelling the existence of distinct agent \textit{types} or \textit{groups}, which they leverage to achieve a more sample-efficient understanding of the agents in their environment. This idea is related to recent work in the area of \textit{social structure learning} concerned with computational accounts of how people learn latent groups from observations of individuals' choices \citep{Gershman2017, Gershman2020}. Given an assumption that members of the same latent group or type share consistently similar reward functions, then the group membership of any particular agent can serve in essence as a compressed representation of their reward function; and thus of their suitability as an imitation target. Future work will explore these ideas in greater depth, considering more specifically the question of imitation on the basis of inferred social groups, through both further behavioral experiments and computational modelling. 

\subsection{Selective imitation in machines}
The results of the current work have implications not only for our understanding of human social learning strategies, but also for how we might design artificial agents that use selective imitation to learn more efficiently in unfamiliar environments. Imitation learning, which has long been a common paradigm for behavior learning in robotics \citep{Osa2018, Argall2009}, typically makes the assumption that there is only ever a single possible demonstrator---an assumption that quickly breaks down outside of only the most controlled environments. Giving machines the ability to actively select suitable imitation targets within rich multi-agent environments could pave the way to robots that are able to better navigate the complexity and uncertainty of the real world. By pointing towards certain priors or inductive biases involved in how people generalize reward function inferences across domains and agents, our initial findings may also have value for the development of more humanlike inverse reinforcement learning algorithms. 

\subsection{Conclusion}
In sum, our results, while preliminary, represent an important step towards understanding how theory of mind abilities such as reward function inference can support sophisticated metacognitive social learning strategies that enable people to acquire adaptive behaviors under various sources of uncertainty.

\bibliographystyle{apacite}

\setlength{\bibleftmargin}{.125in}
\setlength{\bibindent}{-\bibleftmargin}

\bibliography{refs.bib}

\end{document}